\documentstyle[12pt,epsf]{article}

\def\beq{\begin{eqnarray}}    
\def\eeq{\end{eqnarray}}      

\def\Tr{\,\mbox{Tr}\,}                  
\def\pa{\partial}                       
\def\al{\alpha}
\def\be{\beta}

\def\ga{\gamma}

\def\ep{\epsilon}

\def\na{\nabla}

\def\Ga{\Gamma}
\def\De{\Delta}

\begin{document}

\begin{titlepage}

\begin{flushright}
IFT -- P.004/98 \\
\end{flushright}

\hfill hep-ph/9712503

\begin{center}

\vspace{1cm}
{\Large  \bf
The action for the (propagating) torsion  and the limits}
\vskip 2mm
{\Large\bf 
on the torsion}
\vskip 2mm
{\Large\bf 
parameters from present experimental data}

\vskip 7mm
\setcounter{page}1
\renewcommand{\thefootnote}{\arabic{footnote}}
\setcounter{footnote}0

{\bf A.S.Belyaev} $^{\mbox{a,b,}}$\footnote{e-mail:
belyaev@ift.unesp.br},
$\,\,\,${\bf Ilya L.Shapiro} $^{\mbox{c,d,}}$\footnote{e-mail:
shapiro@ibitipoca.fisica.ufjf.br}

\vspace{7mm}

$^{\mbox{a}}$ {\it Instituto de F\'\i sica Te\' orica, Universidade 
Estadual Paulista,\\
Rua Pamplona 145, 01405-900 - S\~ao Paolo, S.P., Brasil}

\vspace{3mm}

$^{\mbox{b}}$ {\it Skobeltsyn Institute of Nuclear Physics,
Moscow State University, \\ 119 899, Moscow, Russian Federation}

\vspace{3mm}

{\sl $^{c}\,\,$
Departamento de Fisica, Universidade Federal de Juiz de Fora,
36036-330, MG -- Brazil }

\vspace{3mm}

{\sl $^{d}\,\,$
Tomsk State Pedagogical University, 634041, Tomsk, Russia }

\vspace{7mm}
\end{center}

\begin{abstract}
Starting from the well established form of the Dirac action
coupled to the electromagnetic and torsion field we find that
there is some additional softly broken local symmetry associated
with torsion. This symmetry fixes the form of divergences of the
effective action after the spinor fields are integrated out.
Then the requirement of renormalizability fixes the torsion
field to be equivalent to some massive pseudovector and its
action is fixed with accuracy to the values of coupling constant
of torsion-spinor interaction, mass of the torsion and higher
derivative terms. Implementing this action into the abelian
sector of the Standard Model we establish the upper bounds
on the torsion mass  and coupling. In our study 
we  used results of present experimental 
limits on four-fermion contact interaction (LEP, HERA, SLAC, SLD, CCFR)
and TEVATRON limits on the cross section  of new gauge boson, 
which could 
be produced as a resonance at high energy $p\bar{p}$ collisions.
\end{abstract}

 \vskip 0.5cm
\end{titlepage}

\vskip 6mm
\noindent
{\large\bf Introduction}
\vskip 2mm

The great success of the Standard Model (SM) in the predictions of
the results of experiments on the accelerators is very impressive.
On the other hand the common point of view is that SM and its direct
generalizations such as GUT's can not serve as a fundamental theory
at a very high energies because they do not include quantum gravity.
These theories should be regarded as an effective field theories
which are only valid at the restricted energy range \cite{weinberg}.
If the fundamental "ultimate" theory will be someday achieved,
it will probably differ from the conventional quantum field theory
and include the nonlocal implications. As an example of such a
theory one can consider the string theory which produce gravity as
an induced interaction and thus solves the quantum gravity problem.
In fact the construction of mathematically consistent
string theory (or its generalizations) is only part of the work
that has to be done. The next problem is to understand in which way 
the unified theory can manifest itself.

One has to notice that the consistent string theory
predicts, along with the metric, other components of the gravitational
field. In particular, the antisymmetric second rank field $A_{\al\be}$
enters the
string effective action via its antisymmetrized derivatives
$\,T_{\al\be\ga} = \pa_{[\al}\,A_{\be\ga]}\,\,$ \cite{GSW} which
are usually referred to as the antisymmetric torsion field.
Therefore the torsion field is predicted by string theory
and it is quite interesting to establish what the effects of torsion
can look like. Recently there were an interesting works devoted to this
problem. In particular, the effects of external background torsion to the
quantized matter fields were discussed in
\cite{bush1,bush2,babush,hammond,hammond2,doma1,doma2,lamme,rytor},
and
some of these papers \cite{hammond2,lamme} contain numerical bounds 
on the possible torsion effects. The phenomenological consequences of the
propagating torsion were considered in \cite{carroll}.
The first of this papers contains an interesting discussions in the
contest of effective field theory and some upper bounds for torsion.
However in this paper the torsion is regarded as a pseudoscalar
longitudinal mode of the antisymmetric tensor while the vector
transversal mode was neglected. The purpose of this
letter is to prove that the action of dynamical torsion necessary
contains massive vector field and to evaluate its possible observational 
consequences. We derive the action for the torsion
pseudovector and find that it contains two free parameters (one of them is 
torsion mass) and study the phenomenological consequences of this action.
As  the result of this study we obtain an upper bounds on the parameters 
of the torsion action using present experimental data.

The paper is organized in the  following way. In the next section we
introduce a basic notations, give a very brief review of gravity with
torsion and establish an additional softly broken symmetry which holds
for the spinor field coupled to torsion (see also \cite{rytor}).
Then we discuss the possible form of divergences which may appear
in the theory with propagating
torsion and find the form of the action which provides the
renormalizability. Section 3 is devoted to the phenomenological
consequences of the propagating torsion, and in section 4 we draw some
conclusions and outline the perspectives for the future study.

\vskip 6mm
\noindent
{\large\bf Spinor field coupled to torsion and electromagnetic field}
\vskip 2mm

Let us give a very brief review for the basical notions of the gravity
with torsion. One can find more detailed introduction to the gravity
with torsion in Refs. \cite{hehl,hehl-review} and to the quantum
field theory in curved space-time with torsion in \cite{book},
which notations we use below.
The metric $g_{\mu\nu}$ and torsion $T^\alpha_{\;\beta\gamma}$
have to be considered as an independent characteristics 
of the space - time.
Since here we are interested in the torsion effects only,
the metric is supposed to be flat Minkowski one everywhere.
In the theory with  torsion the covariant derivative $\tilde{\nabla}$
is based on the nonsymmetric connection
$\tilde{\Gamma}^\alpha_{\;\beta\gamma}$ with
\beq
\tilde{\Gamma}^\alpha_{\;\beta\gamma} -
\tilde{\Gamma}^\alpha_{\;\gamma\beta} =
T^\alpha_{\;\beta\gamma},
\label{tor}
\eeq
Indeed $\tilde{\Gamma}^\alpha_{\;\beta\gamma}$ is not a tensor
because one can use the curvilinear coordinates.
The  metricity condition
$\,\tilde{\nabla}_\mu g_{\alpha\beta} = 0\,$ enables one to express
 the connection through metric and torsion in a unique way as
\beq
\tilde{\Gamma}^\alpha_{\;\beta\gamma} = \left\{^{\,\al}_{\be\ga}\right\}+
\frac{1}{2}\left( T^\alpha_{\;\beta\gamma} -
T^{\;\alpha}_{\beta\cdot\gamma} - T^{\;\alpha}_{\gamma\cdot\beta} \right)
\label{con}
\eeq
where
$\left\{^{\,\al}_{\be\ga}\right\}$
is the Christoffel symbol. It is important that the rest of the formula
(\ref{con}) is tensor and thus it can not be eliminated by the
change of coordinates.
It proves convenient to divide the torsion field into three irreducible
components:
Trace $T_{\beta} = T^\alpha_{\,\;\beta\alpha}$;
Pseudotrace
$\;S^{\nu} = \epsilon^{\alpha\beta\mu\nu}T_{\alpha\beta\mu}\;$ and
the tensor
$\;q^\alpha_{\;\beta\gamma}\;$, for which the two conditions are satisfied
$$
\;q^\alpha_{\;\beta\alpha} = 0;\;\;\;\;\;
\epsilon^{\alpha\beta\mu\nu}q_{\alpha\beta\mu} =0.\;\;
$$

In the string-induced action, which depends on the completely
antisymmetric torsion, only the pseudovector part
$\,S_\mu\,$ is present, and thus one can
always set
\beq
T_{\alpha\beta\mu} =
\frac{1}{6}\, \varepsilon_{\alpha\beta\mu\nu}\,S^{\nu}.
\label{anti}
\eeq
Below we shall use the pseudovector $S_\mu$ as parametrization for the
antisymmetric torsion tensor.

The minimal action of the Dirac spinor fields in an external
gravitational field with torsion follows from the standard procedure
(see, for example, \cite{book}).
One has to change the partial derivatives $\pa_\mu$
to the covariant ones $\tilde{\nabla}_\mu$. In case of the
general Riemann-Cartan manifold \cite{hehl,book} one has also to
change the flat metric $\eta^{\mu\nu}$ to the general one
$g^{\mu\nu}$ and generalize the
volume element $d^4x$ to the covariant one $d^4x\sqrt{-g}$.
In our case $g_{\mu\nu}=\eta_{\mu\nu}$ this
procedure leads to the expression:
$$
S_{\frac12 , min} = \frac{i}{2}\,\int d^4x\,\left(\,
{\bar \psi}\,
\ga^\al \,\tilde{\na}_\al\psi -
 \tilde{\na}_\al{\bar \psi}
\,\ga^\al \,\psi -2im {\bar \psi} \psi\,  \right)
$$
\beq
= i\,\int d^4x \, {\bar \psi} \left(\,
\ga^\al \,{\pa}_\al - \frac{i}{8}\,\ga_5\ga^\al\,S_\al - im\, \right)\psi
\label{dirac}
\eeq
Here ${\bar \na}$ is covariant derivative of the spinor field
(see \cite{book} for the details.)

The renormalizability of the gauge model in an external torsion field
requires the nonminimal interaction of the
spinor and scalar fields with torsion \cite{bush1}. Later on we shall see
that this is reasonable to introduce the nonminimal spinor-torsion
interaction 
in the theory with the propagating
torsion as well. For this reason we shall start our study from the
action of the Dirac spinor
nonminimally coupled with the electromagnetic and torsion fields
\beq
S_{1/2}= i\,\int d^4x\sqrt{-g}\,{\bar \psi}\, \left[
\ga^\al \,\left( {\pa}_\al - ieA_\al + i\eta\ga_5\,S_\al\,\right)
- im \right]\,\psi
\label{diraconly}
\eeq

The new interaction with torsion doesn't spoil
the invariance of the above action under usual gauge transformation:
\beq
\psi' = \psi\,e^{\al(x)}
,\,\,\,\,\,\,\,\,\,\,\,\,\,\,
{\bar {\psi}}' = {\bar {\psi}}\,e^{- \al(x)}
,\,\,\,\,\,\,\,\,\,\,\,\,\,\,
A_\mu ' = A_\mu - {e}^{-1}\, \pa_\mu\al(x)
\label{trans1}
\eeq
Moreover the massless  part of  the action (\ref{diraconly})
is invariant under the transformation in which the pseudotrace of
torsion plays the role of the gauge field
\beq
 \psi' = \psi\,e^{\ga_5\be(x)}
,\,\,\,\,\,\,\,\,\,\,\,\,\,\,
{\bar {\psi}}' = {\bar {\psi}}\,e^{\ga_5\be(x)}
,\,\,\,\,\,\,\,\,\,\,\,\,\,\,
S_\mu ' = S_\mu - {\eta}^{-1}\, \pa_\mu\be(x)
\label{trans}
\eeq
Thus in the massless sector of the theory one faces generalized
gauge symmetry depending on scalar $\al(x)$ and pseudoscalar
 $\be(x)$ parameters of transformation. As it will be shown below,
this new symmetry requires torsion to be massive vector field
and furthermore the action of torsion is fixed with
accuracy to the values of nonminimal parameter $\eta$, mass of the
torsion $M_{ts}$ and possible higher derivative terms.

In the framework of effective field theory the effects of a very
massive fields are suppressed by the factors of $\,\mu^2 / M^2$ where
$M$ is the mass of the field and $\mu$ the typical energy of the
process. Thus if we take $M_{ts}$ to be of the Planck order then the
effects of torsion will be negligible at the energies available at the
modern experimental facilities. The hypothesis of torsion, propagation
at energies lower than the Planck one
supposes that $M$ is essentially smaller than the Planck mass. Then
we have two options: take torsion to be massless or consider the mass
of torsion as a free parameter which should be defined on the
experimental basis. As far as the torsion can propagate, one has to
incorporate it into the SM along with other vector fields. 
Let us discuss the form of the action for
torsion, which leads to the consistent quantum theory. The higher
derivative terms in the action, in general,
lead to the unphysical ghosts and to the consequent violation of
unitarity. Therefore we restrict the torsion action by the second
derivative and zero-derivative terms.
The general action including these terms has the following form:
\beq
S_{tor} = \int d^4\,\left\{\, -a
S_{\mu\nu}S^{\mu\nu} + b(\pa_\mu S^\mu)^2
+ M_{ts}^2\,S_\mu S^\mu\,\right\}
\label{geral}
\eeq
where $\,S_{\mu\nu} = \pa_\mu S_\nu - \pa_\mu S_\nu\,$ and $a,b$
are some positive parameters. The action (\ref{geral}) contains both
transversal vector mode and the longitudinal model which is in fact
equivalent to the scalar\footnote{This kind of torsion equivalent to the 
pseudoscalar field was introduced in \cite{novello} in order to 
maintain the gauge invariance of abelian vector 
field in the Riemann-Cartan spacetime.} 
(see, for example, discussion in \cite{carroll})
In particular, in the $a=0$ case only the scalar mode,
and for $b=0$ only the vector mode propagate.
It is well known~\cite{vector} that in the unitary
theory of the vector field both longitudinal and transversal
modes can not propagate, and therefore, in order to have consistent
theory of torsion one has to choose one of parameter $a,b\,$ to be zero
\footnote{We remark that earlier the unitarity was effectively used for 
the construction of the action of gravity with torsion in 
\cite{nevill,seznie}.}.

In fact the only correct choice is $b=0$. To see this one has to reveal
that the symmetry, which is spoiled by the massive terms only, is
always preserved in the renormalization of the dimensionless couplings
constants of the theory. In other words, the divergences and
corresponding local counterterms, which produce the dimensionless
renormalization constants, do not depend on the dimensional parameters
such as the masses of the fields. The symmetry (\ref{trans}) holds for
the massless part of the action (\ref{diraconly}) and therefore
on the general grounds one has to expect that the gauge invariant
counterterm $\,\,\int S_{\mu\nu}^2\,\,$ appears if we take the loop
corrections into account.

We want emphasize that in the framework of effective field
theory the level of approximation for taking into
account the massive fields is qualitatively the same 
for the tree level and 
for the lower loop effects. Therefore as far as the propagating 
torsion is considered and the kinetic term in (\ref{geral}) 
is taken into account, one has to formulate the theory
as renormalizable. Neglecting the high energy effects while the
low energy amplitudes are considered may mean that we disregard
some higher derivative terms. However the violation of the 
renormalizability
in that sectors of the theory which are taken into account is impossible.
For instance, if we start from the purely scalar longitudinal torsion
(as the authors of \cite{carroll} did) then the transversal term
$\,\,\int S_{\mu\nu}^2\,\,$ will arise with the divergent coefficient
and this will indeed violate both the finiteness of the effective
action and the unitarity of the $S$-matrix. All this is true even
in the case that only the tree-level effects are evaluated, if only such
consideration is regarded as an approximation to any reasonable
quantum theory.

Thus the kinetic term of the torsion action is given by the
Eq. (\ref{geral}) with $b=0$. As concerns the massive term it is not
forbidden by the symmetry (\ref{trans}), because the last is softly
broken. Therefore apriory there are no reasons to suppose that $M_{ts}=0$.
The only one question is: whether the massive counterterm really appears
if we take into account the fermion loops. To investigate this
we have performed the one-loop calculation of divergences in the
theory (\ref{diraconly}), using the standard Schwinger-deWitt technique
and dimensional regularization (one can see \cite{book} for introduction
and references). The result of these calculations of
\footnote{The same calculation for the massless theory in curved
space-time has been performed in \cite{buodsh}
(see also references there and in hep-th version of \cite{babush}).
More details about (\ref{contra}) and consequent 
renormalization group equations will be given elsewhere \cite{futuro}. }
\beq
\Ga_{div} [A, S] = - \Tr\ln \left[i
\ga^\al \,\left( {\pa}_\al - ieA_\al + i\eta\ga_5\,S_\al\,\right)
- im \right]_{\,div}
\label{efac}
\eeq
Is the following counterterm:
\beq
\De S [A_\mu, S\al] = \frac{1}{\varepsilon} \int d^4x\,\left\{
\frac{2e^2}{3}F_{\mu\nu}F^{\mu\nu} +
\frac{2\eta^2}{3}S_{\mu\nu}S^{\mu\nu}
- \frac{ie\eta}{3}\ep^{\al\be\mu\nu}S_{\mu\nu}F_{\al\be}
+ 8m^2\eta^2S^\mu S_\mu  \right\}
\label{contra}
\eeq
with $\,\varepsilon = (4\pi)^2\,(n-4).\,$
Here we have neglected all surface terms except
$\ep^{\al\be\mu\nu}\,S_{\mu\nu}F_{\al\be}$, because it can,
in principle,
lead to quantum anomaly. The phenomenological consideration
below is restricted by the tree-level effects and therefore this
term is beyond the scope of our present interest.
The form of the counterterms (\ref{contra}) indicates that the massive
term in the action of torsion is indeed necessary for the
renormalizability and hence the correct form of the action is
\beq
S_{tor} = \int d^4\,\left\{\, -\frac14\,S_{\mu\nu}S^{\mu\nu}
+ M_{ts}^2\, S_\mu S^\mu\,\right\}
\label{action}
\eeq
In the last expression we put the conventional coefficient $\,-1/4$
in front of the kinetic term. With respect to the renormalization
this means that we (in a direct analogy
with QED) can remove the kinetic counterterm by the renormalization
of the field $S_\mu$ and then renormalize the parameter $\eta$ in
the action (\ref{diraconly}) such that the combination
$\eta S_\mu$ is the same for the bare and renormalized quantities.
Instead one can include $1/\eta^2$ into the kinetic term of
(\ref{action}), that should lead to the direct renormalization of
this parameter while the interaction of torsion with spinor has
minimal form (\ref{dirac}) and $S_\mu$ is not renormalized.
Therefore in the case of propagating torsion the difference
between minimal and nonminimal interactions is only the question of
notations on both classical and quantum levels.

In the next section we shall discuss the possible consequences of the 
torsion action at low energies and find some numerical upper bounds
for torsion. From the string theory point of view $M_{ts}$
should have the value of the Planck order. Indeed this choice doesn't give
a chance to make any speculations and estimates using an available
experimental data because they are obtained at the energies which are 
(at least) 16 orders smaller than the Planck ones. Therefore we 
suppose that as a result of some cancelation the string inspired mass
$M_{ts}$ vanishes and then take $M_{ts}$ to be some
free parameter of the theory.
We shall consider two different possibilities: i) torsion is much more 
heavy than other particles of SM and $\,\,$ 
ii) torsion has a mass comparable to that
of other particles. In the last case one meets a propagating 
particle which 
must be treated on an equal footing with other constituents of SM.
Contrary to that, the very heavy torsion leads to the effective contact 
four-fermion interactions. 

Consider this in some more details, starting 
from the actions (\ref{diraconly}) and (\ref{action}). Since 
the massive term dominates over the covariant kinetic 
part of the action, the last can be disregarded. Then the action
$S_{1/2} + S_{tor}$ leads to the algebraic equation of motion 
for $\,S_\mu$. The solution of this equation can be substituted back to 
$\,S_{1/2} + S_{tor}\,$ and thus produce the contact four-fermion 
interaction term
\beq
{\cal L}_{int} = - \frac{\eta^2}{M_{ts}^2}\,
({\bar \psi}\ga_5\ga^\mu\psi)\,({\bar \psi}\ga_5\ga_\mu\psi)
\label{contact}
\eeq

As one can see the only one quantity which appears in this approach is the 
ratio ${M_{ts}}/{\eta}\,$ 
and therefore the phenomenological consequenses may depend only on 
single parameter. 

In the next 
section we consider the upper bounds
on the phenomenological manifestations of contact 
interactions between quarks and leptons and also some observational
limits for the propagating torsion.  


\vskip 6mm
\noindent
{\large\bf 3. Possible physical observables related with torsion action
and the limits on the torsion parameters}
\vskip 2mm

In this section we put the limits on the parameters of the torsion action
using results of various experiments\footnote{The detailed exposition 
will be presented elsewhere \cite{futuro}}.
Torsion being a pseudo-vector particle interacting with fermions
might give therefore different physical observables.

Physical observables related with torsion depend on the two basic 
parameters, namely on the
torsion mass $M_{ts}$ and the constant of the interaction 
between torsion and fermion fields $\eta$. 
In the course of our study we choose, for the sake of simplicity,
all the torsion couplings with fermions to be the same $\eta$ 
\cite{futuro}. 
This enables one to put the limits in the two
dimensional ($M_{ts}$-$\eta$) parameter space using 
the present experimental data.

The straightforward consequence of the action term  for torsion
interaction with fermion field is the effective four-fermion contact
interaction of  leptons and quarks (\ref{contact}). 
Four-fermion interaction effectively appears 
for the torsion with  mass much higher than the energy
scale available at present colliders. There are several experiments from 
which the constraints on the contact four-fermion interactions come:

\vskip 1mm
\noindent
1)Experiments on polarized electron-nucleus scattering -
SLAC e-D scattering experiment\cite{slac}, Mainz e-Be scattering
experiment \cite{mainz} and bates e-C scattering experiment \cite{bates};
\vskip 0.8mm
\noindent
2)Atomic physics parity violations measures \cite{apv}
electron-quark coupling that are different from those tested at 
high energy experiment
 provides alternative constraints on new physics.
\vskip 0.8mm
\noindent
3) $e^+e^-$ experiments  - SLD, LEP1, LEP1.5 and LEP2 (see for example
\cite{lep,opal,l3,aleph,lang});
\vskip 0.8mm
\noindent
4)Neutrino-Nucleon DIS experiments -- CCFR collaboration obtained model
independent constraint on the
effective $\nu\nu q q$ coupling \cite{ccfr};
\vskip 0.8mm
\noindent
5)HERA experiment for the polarized
lepton-proton beams interactions~\cite{hera}.
\vskip 1mm

First we consider a limits on the contact interactions induced by 
torsion. The contact four-fermion interaction may be described by the
Lagrangian~\cite{eich} of the most general form:
\begin{equation}
L_{\psi'\psi'\psi\psi}=g^2\sum_{i,j=L,R}\sum_{q=u,d}\frac{\epsilon_{ij}} 
{(\Lambda_{ij}^{\epsilon})^2}
(\bar \psi'_i\gamma_\mu \psi'_i)(\bar \psi_j\gamma^\mu \psi_j)
\label{cont}
\end{equation}
Subscrips i,j refer to different fermion helicities:
$\, \psi^{(')}_i= \psi^{(')}_{R,L}= (1\pm\gamma_5)/2\cdot \psi^{(')}$; 
where $ \psi^{(')}$ could be quark or lepton; $\,\,\Lambda_{ij}\,$ 
represents the mass scale of the exchanged new particle; coupling strength
is fixed by the relation: $\,g^2/4\pi=1$, 
the sign factor  $\,\epsilon_{ij}=\pm 1\,$ 
allows for either constructive or destructive interference with 
Standard Model (SM) $\gamma$ and $Z$-boson exchange amplitudes.
The formula (\ref{cont}) can be successfully used for the study of the 
torsion-induced contact interactions because it includes an axial-axial
current interactions as a particular case. 

Recently the global study of the
electron-electron-quark-quark($eeqq$) interaction sector of the 
SM~\cite{global} have been done using data from all mentioned 
above experiments (except the fifth item). 
The limits established in this paper 
are the best in comparison with the previous ones.
Since the effective contact Lagrangian for torsion has axial-axial 
structure we used the limits obtained in paper~\cite{global} for this 
kind of interaction.
We remark that 
the limits on $\Lambda$ of paper~\cite{hera} are quite  close to those 
of ~\cite{global}
but are not suitable for our analysis because  axial-axial 
interaction have not been studied in ~\cite{hera}.
Axial-axial current may be expressed through RR,LR,RL and LR currents in the
following way:
 \begin{equation}
j_\mu^A j_\mu^A=
\frac{j_\mu^L j_\mu^L+j_\mu^R j_\mu^R-j_\mu^L j_\mu^R-j_\mu^R j_\mu^L}{4}\,.
\label{aa}
\end{equation}
For the axial-axial 
$\,eeqq\,$ interactions~(\ref{cont}) takes the form (we put $g^2=4\pi$) :
\begin{equation}
L_{eeqq}=-\frac{4\pi}{{(\Lambda_{AA}^{\epsilon})^2}}
(\bar e\gamma_\mu\gamma_5 e)(\bar q\gamma^\mu \gamma_5 q)
\label{contaa}
\end{equation}
The limit for the contact axial-axial $eeqq$ interactions comes
from the global analysis of Ref.~\cite{global}:
\begin{equation}
\frac{4\pi}{\Lambda_{AA}^2} < 0.36\mbox{ TeV} ^{-2}
\label{glob}
\end{equation}
For the parameters of the effective contact four-fermion 
interactions of general form
(13) and contact four fermion interactions induced by 
torsion (12) we have the
following relations:
\begin{equation}
\frac{\eta^2}{M_{ts}^2}=\frac{4\pi}{{\Lambda_{AA}}^2}
\label{rel}
\end{equation}
>From (\ref{glob}) and (\ref{aa}) one gets the following limit
on torsion parameters:
\begin{equation}
\frac{\eta}{M_{ts}}<0.6\mbox{ TeV}^{-1}\; \Rightarrow \;
M_{ts}>1.7\mbox{ TeV }\cdot\eta
\label{globfin}
\end{equation}
\begin{figure}[htb]
   \begin{center}
    \vskip -0.8cm\hspace*{-0.5cm}
    \epsfxsize=8cm\epsffile{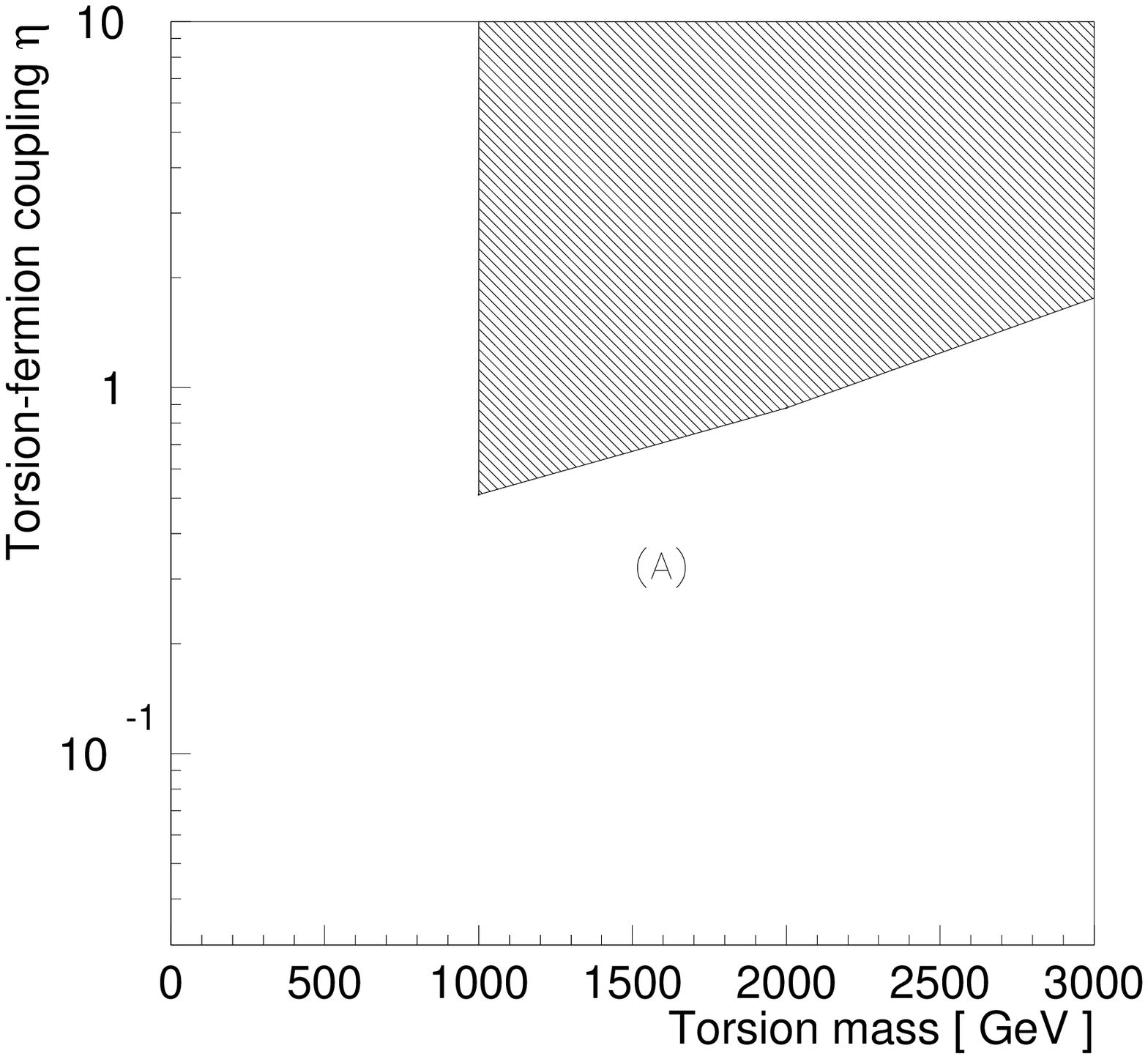}\epsfxsize=8cm\epsffile{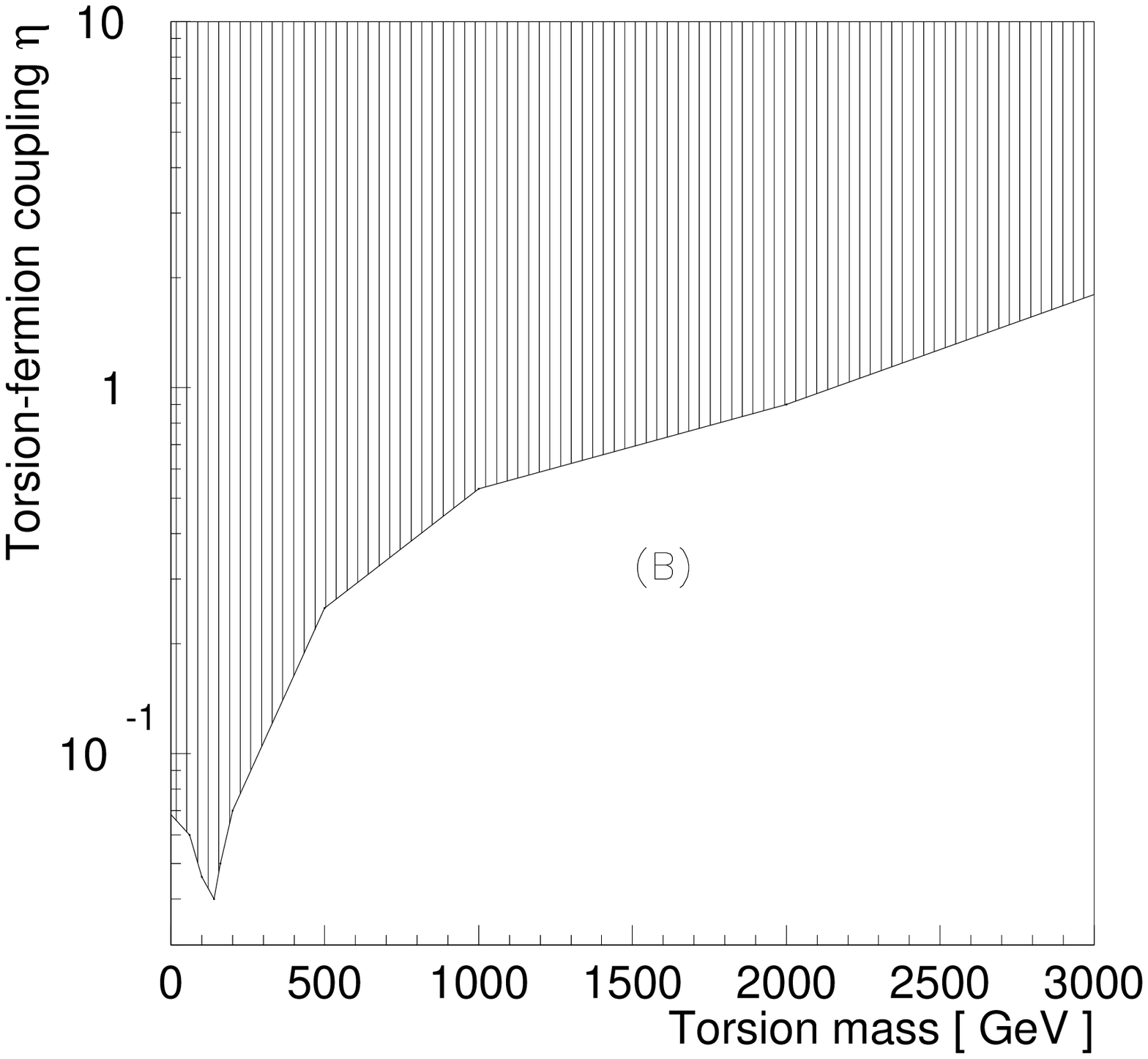}
    \vskip -0.8cm\hspace*{-0.5cm}
    \epsfxsize=8cm\epsffile{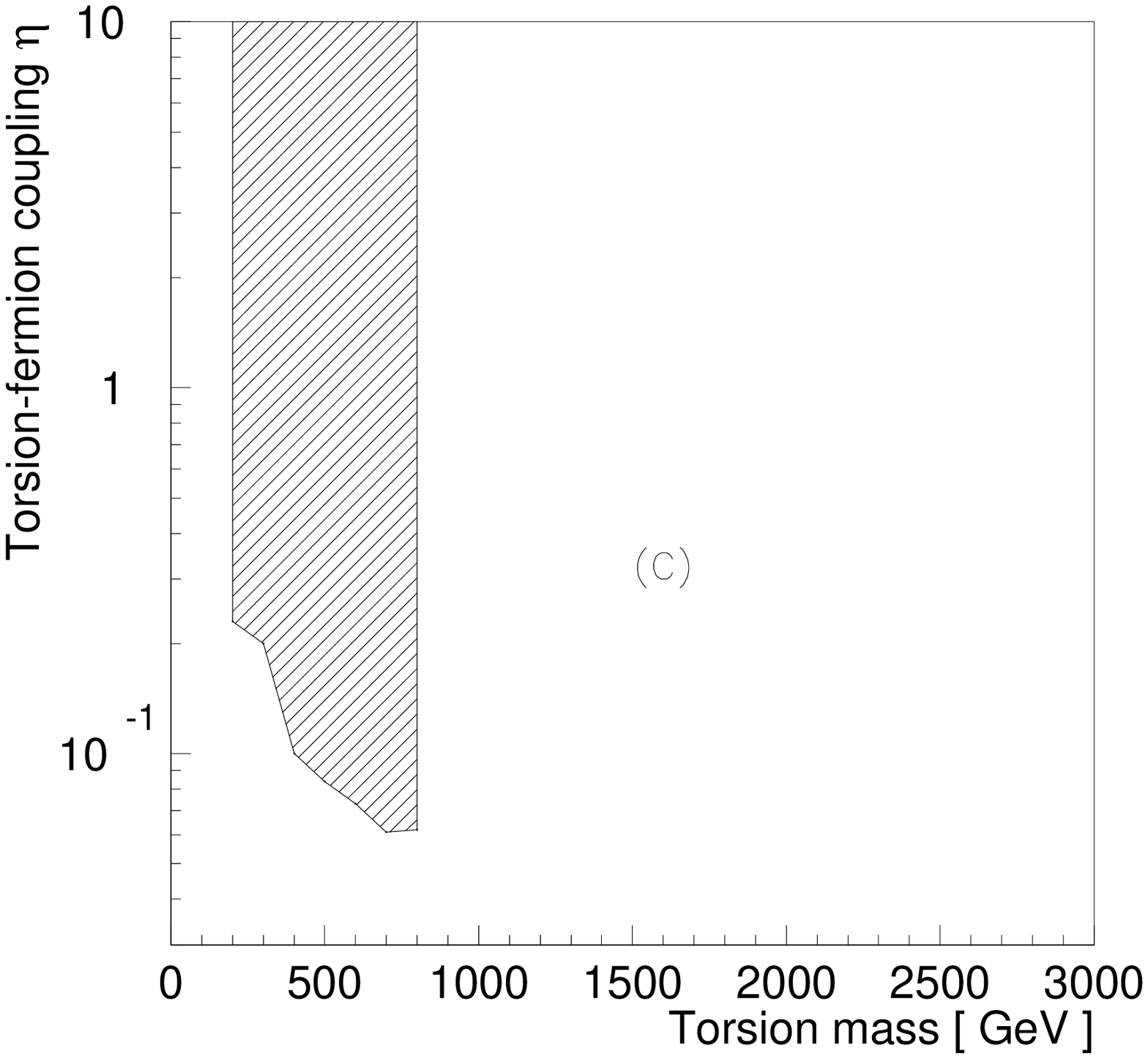}\epsfxsize=8cm\epsffile{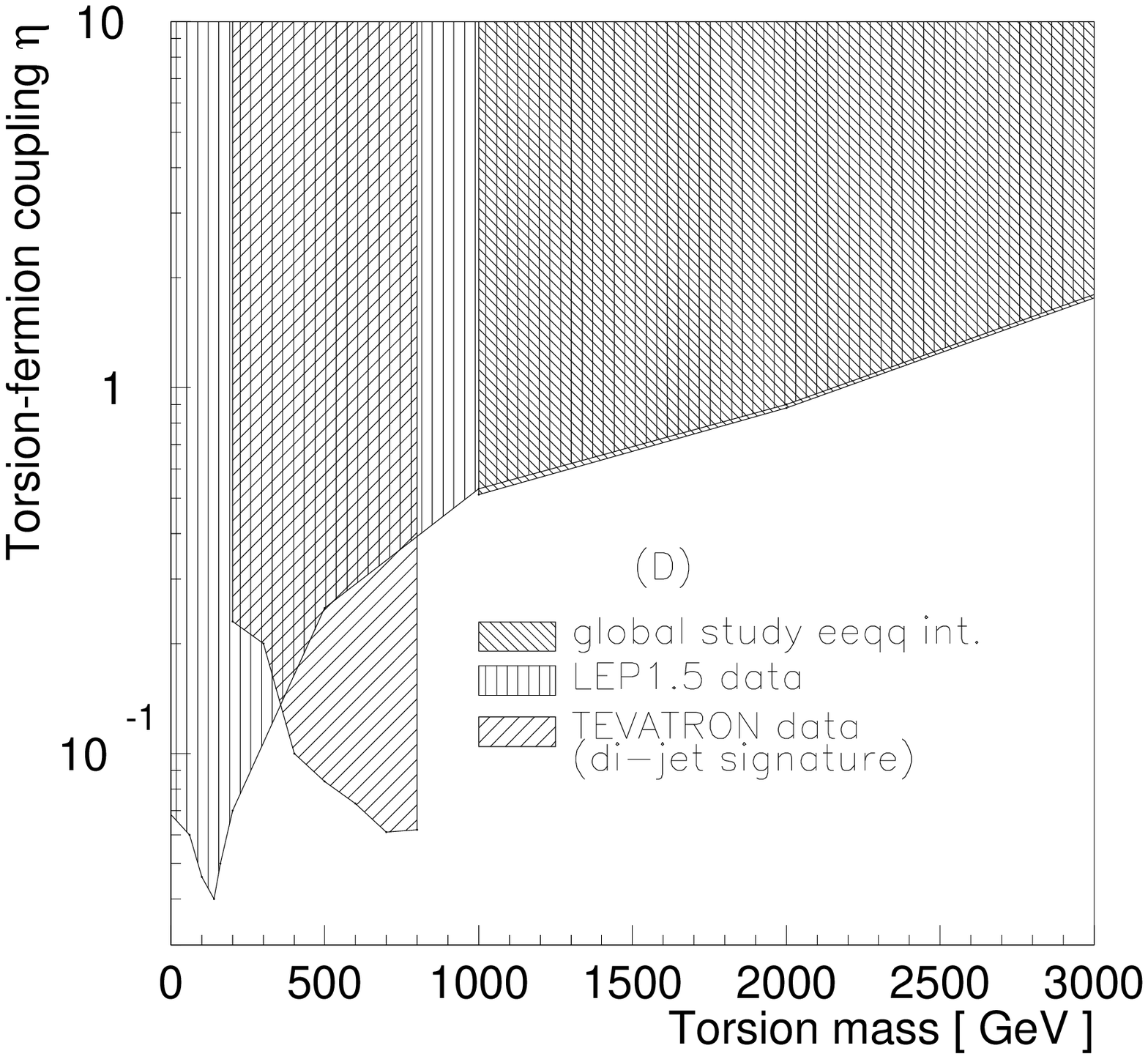}\\
  \end{center}
  \vspace*{-1.0cm}
    \caption{Allowed regions for $M_{ts}$ and  $\eta$ coming from
    global study of electron-quark contact interactions (A),
    LEP1.5 (B)  and TEVETRON data (C).  
    (D) -- combined limit. Hatched region is excluded by   experiments 
mentioned above}
\end{figure}
The limits on $M_{ts}$ and $\eta$ coming from the (\ref{globfin})
is shown in Figure~1(A). Some remark about the energy limits 
taken in this plot is in order.
We started exclusion region from $M_{ts}=$~1~TeV. This choice is 
related with the fact that the
application of effective-contact interactions (\ref{contact}) 
is valid up to the
certain mass of the torsion below which an exact calculation 
(regarding the field $S_\mu$ as dynamical) should be done.
The relative data of the two approaches are shown on the Figure 2, 
where the results for gauge interaction~(\ref{action}) 
and contact interactions~(\ref{cont}) for torsion are compared. 
As an example  we have calculated total cross section for LEP1.5
with $\sqrt{s}=140$ GeV and $\eta$ equal to 0.5. 
One can clearly see that for torsion heavier than 1 TeV the approximation 
of the effective contact interaction works almost perfectly, reproducing 
the result for the
exact calculation with 0.1\% accuracy. Therefore the scale 1 TeV 
is appropriate starting point for  putting the limit on 
torsion parameters using the Lagrangian with contact interactions. 
\begin{figure}[htb]
  \vspace*{-1.0cm}
    \epsfxsize=12cm
    \epsffile{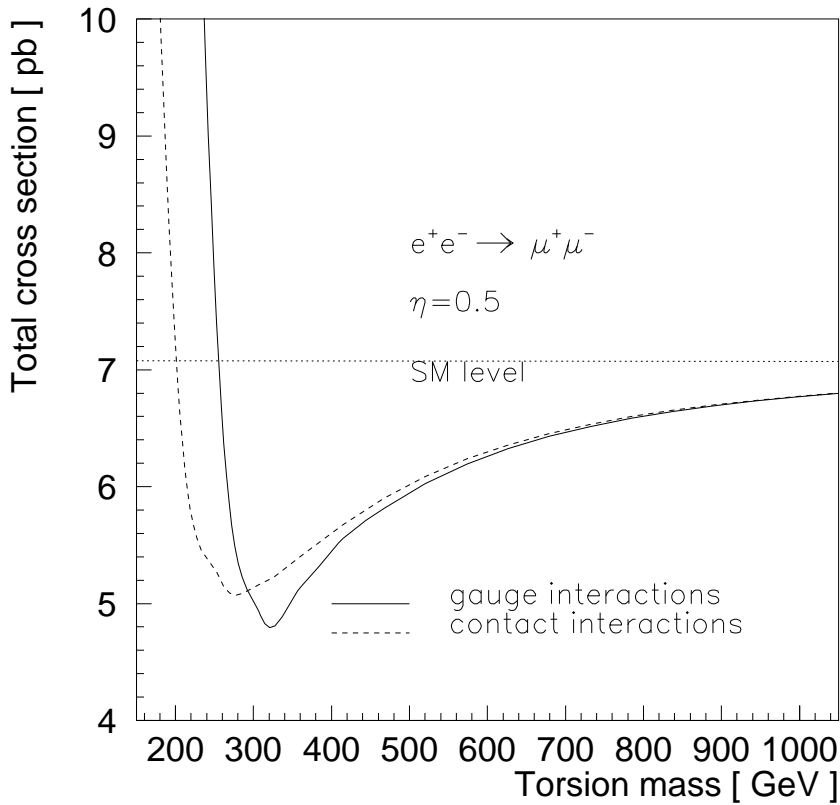}
 \vspace*{-1.0cm}
    \caption{Comparison  of the total cross sections
    of $e^+e^0\rightarrow \mu^+\mu^-$ process 
    for gauge and contact interactions }
 \end{figure}
Scenario with light torsion is in general more difficult because here
we have two independent parameters and thus are enforced to 
study the $2$-dimensional restrictions from the experimental data.
Indeed there is no rigid border between two cases, as we shall see below.

For  constraining $M_{ts}-\eta$ parameter space for
light torsion we use results of  LEP1.5 analysis of paper\cite{opal}:
the cross section of $e^+e^-\rightarrow e^+e^-(\mu^+\mu^-)$ process 
was measured with 
accuracy 1-2\%. We used this fact  to put the limits on torsion mass 
and coupling:
90\% acceptance for electron and 60\% for muon channels was 
assumed, the total 
cross section for these reaction 
were calculated and 4\% deviation from the Standard
Model prediction was taken for establishing the limits.
The results constraints is shown in Figure~1(B).

 The torsion with the mass
in the range of  present colliders could be produced in fermion-fermion 
interactions as a resonance, decaying to fermion pair.
The most promising collider for search the signature 
of such type is TEVATRON.
Search for New Particles Decaying to two-jets 
has been done recently by D0 and CDF collaborations \cite{d0}.
The data that we use in our analysis are extracted from  figure presented 
D0 collaboration which established the limit on the production 
cross section
of $Z'$ and $W'$ bosons .
Here we assume also 90\% events efficiency 
(including efficiency of kinematical cuts
and trigger efficiency)
 and calculated the cross
section for torsion production at TEVATRON. Then we applied   
D0 limit at 95\% CL
 for torsion production cross section and converted into  the
limit for  $M_{ts}-\eta$ plane. This limit is  shown in figure 1~(C).
The points for the exclusion curve are given in Table~1. 

\begin{table}[htb]
\begin{center}
\begin{tabular}{ l | l | l | l | l | l | l | l }
$M_{ts}$(GeV)& 200&300 &400 & 500 &600  & 700 & 800\\
\hline
$\eta$     &0.23&0.20&0.10&0.084&0.073&0.061&0.062
\end{tabular}
\end{center}
\caption{Points for exclusion curve in $M_{ts}-\eta$ plane 
from TEVATRON data}
\end{table}
One can see that the limits on $\eta$ coming from these analysis
are much better in comparison with those from the LEP data.
Combined exclusion plot for $M_{ts}-\eta$ plane is presented in Fig.~1(D).

It should be stressed that all numerical and symbolic calculations 
for establishing limits on torsion parameter space have been done using
CompHEP software package \cite{comp} where the torsion action 
was introduced.

\section*{4. Conclusions}

We derived the action of propagating torsion and implemented
it into the abelian sector of the Standard Model. It was shown that the
only one action of torsion which leads to consistent (unitary and 
renormalizable) theory 
includes propagating pseudovector massive particle with softly broken 
(new) gauge symmetry. 
Starting from this action for torsion, we have established some upper 
bounds on the torsion mass and torsion-spinor coupling constant 
(which is supposed to be universal) using combined limit for four-fermion 
interactions, LEP and TEVATRON data. 
For  heavy torsion the limit is described by
relation~(\ref{globfin}) while for light torsion with the mass below 1 TeV
limits coming from LEP and TEVATRON data  bound $\eta$ to be less than 
0.1-0.02 depending on $M_{ts}$. We are going to give more details about 
these results in future publication \cite{futuro}. 

Another interesting aspect concerns the possibility to implement scalar
fields. The study of the GUT-like 
theories with Yukawa scalar-spinor interactions in 
an external torsion field
\cite{bush1} has demonstrated the necessity to introduce the nonminimal 
interaction of scalar fields with torsion. When one is considering the 
propagating torsion, all the diagrams which were considered in \cite{bush1}
still give the same contributions and therefore it is very probable
that the full consistent theory contains a nonminimal scalar-torsion 
interaction and also the torsion self-interaction term. 
These interaction will not change  significantly 
observables  considered  in this paper since the physical effects from 
scalar-torsion interaction and also 
the torsion self-interaction will only appear at loop level.
But they might lead to some other physical observables which probably
could allow us to improve the limits on the torsion parameters.
Moreover some formal questions related with the torsion 
self-interaction can be addressed. We hope to be back to these problems in 
the close future. 

\vskip 5mm
\noindent
{\bf Acknowledgments}
\vskip 2mm

One of the authors (I.L.Sh.) is grateful 
to M. Asorey, I.L. Buchbinder, I.B. Khriplovich and T. Kinoshita for
stimulating discussions. He also acknowledges 
warm hospitality of Departamento de Fisica, Universidade
Federal de Juiz de Fora and partial support by Russian Foundation for Basic
Research under the project No.96-02-16017.
A.S.B. acknowledges support from 
Funda\c{c}\~ao de Amparo \`a Pesquisa do Estado de S\~ao Paulo
(FAPESP).

\end{document}